\documentclass[12pt,psfig,a4,]{article}
\usepackage{geometry}
\usepackage{graphics,algorithmicx,algpseudocode,algorithm,pseudocode}
\usepackage{graphicx,caption,subcaption}
\usepackage{hyperref}
\usepackage{setspace}
\usepackage{amsmath,amssymb,amsthm,bbding}
\newcommand{\Lyx}{L\kern-.1667em\lower.25em\hbox{y}\kern-.125emX\spacefactor1000}

\def\easytfoursymbol#1{\mathord{\mathchoice
  {\mbox{\fontsize\tf@size\z@\usefont{T4}{\rmdefault}{m}{it}\char#1}}
  {\mbox{\fontsize\tf@size\z@\usefont{T4}{\rmdefault}{m}{it}\char#1}}
  {\mbox{\fontsize\sf@size\z@\usefont{T4}{\rmdefault}{m}{it}\char#1}}
  {\mbox{\fontsize\ssf@size\z@\usefont{T4}{\rmdefault}{m}{it}\char#1}}
}}

\newcommand{\BigO}[1]{\ensuremath{\operatorname{O}\bigl(#1\bigr)}}
\makeatother

\newcommand{\sg}{\texttt{SortedGreedy[m]}}
\newcommand{\sgt}{\texttt{SortedGreedy[2]}}
\newcommand{\gr}{\texttt{Greedy[m]}}
\newcommand{\grt}{\texttt{Greedy[2]}}
\date{}
\theoremstyle{plain}
\begin{document}
\bibliographystyle{amsplain} 
\pagestyle{plain} 
\pagenumbering{arabic}

\title{Balanced offline allocation of weighted balls into bins}
\author{ \"Omer Demirel, Ivo F.~Sbalzarini  \\
  \multicolumn{1}{p{.7\textwidth}}{\centering\emph{MOSAIC Group, Max Planck Institute of Molecular Cell Biology and Genetics, Center of Systems Biology \\
Pfotenhauerstr.~108, D--01307 Dresden, Germany.}}}
\maketitle


\begin{abstract} 

We propose a sorting-based greedy algorithm called \sg{} for approximately solving the offline version of the $d$-choice weighted balls-into-bins problem where the number of choices for each ball is equal to the number of bins. We assume the ball weights to be non-negative. We compare the performance of the sorting-based algorithm with a na\"{i}ve algorithm called \gr{}. We show that by sorting the input data according to the weights we are able to achieve an order of magnitude smaller \textit{gap} (the weight difference between the heaviest and the lightest bin) for small problems ($\leq4000$ balls), and at least two orders of magnitude smaller \textit{gap} for larger problems. In practice, \sg{} runs almost as fast as \gr{}. This makes sorting-based algorithms favorable for solving offline weighted balls-into-bins problems. \\

\end{abstract} 


\section{Introduction}
The classical \textit{balls-into-bins} problem~\cite{Johnson:1977,Kolchin:1978} considers the sequential placement of $n$ balls into $m$ bins such that the bins are maximally balanced. Historically, the problem is categorized by the types of balls (e.g., uniform~\cite{Azar:1994,Azar:1999} vs.~weighted~\cite{Berenbrink:2008,Talwar:2007,Peres:2010,Dutta:2011}), by the number of bins a ball can choose from (e.g., single-choice vs.~multi-choice~ \cite{Mitzenmacher:2001}), and by the number of balls (e.g., $n=m$~\cite{Raab:1998} vs.~$n>m$ or $n\gg m$~\cite{Berenbrink:2006}). In applications such as load balancing, hashing, and occupancy problems in distributed computing~\cite{Berenbrink:2008,Berenbrink:2006,Czumaj:2003} the $d$-choice variant and its subproblem, the two-choice variant have been the main focus. This is because the mathematical analysis of load-balancing algorithms frequently involves solving a \textit{balls-into-bins} problem where the balls and bins represent tasks and compute units, respectively. \\

Of special practical importance is the weighted case where the balls have individually different weights. Talwar and Wieder~\cite{Talwar:2007} have shown that as long as the weight distribution has finite second moment, the weight difference between the heaviest and the average bin (i.e., the \textit{gap}) is independent of $n$. Peres \textit{et al.}~\cite{Peres:2010} introduced the $(1+\beta)$-choice process analysis, and for $\beta=1$ the \textit{gap} has a bound $\Theta(\log \log n)$ even for the case of weighted balls. Dutta \textit{et al.}~\cite{Dutta:2011} introduced the $IDEA$ algorithm, which provides a constant \textit{gap} with high probability (w.h.p.) even in the heavily loaded case $n\gg m$ in case of an expected constant number of retries or rounds per ball. \\
 
The \textit{offline} version of the weighted balls-into-bins problem has received less attention than the \textit{online} version. We believe, however, that the offline version is as important in practice as the online version, since in compute systems with a priorly known number of tasks the optimal assignment of these tasks to processors and the expected load imbalance can be analyzed by solving an offline balls-into-bins problem. In the offline setting, we define the \textit{gap} as the weight difference between the heaviest and the lightest bin. We do not restrict the distribution from which the balls sample their weights. For simplicity, we assume that a ball can be placed into any bin, thus $d=m$. We propose to initially sort the balls according to their weights and then use a greedy algorithm to place the next heaviest ball into the lightest bin. We show that even for moderate problem sizes ($n<4000$ balls) this sorting-based greedy algorithm results in a 10 to 60-fold smaller \textit{gap} than the na\"{i}ve \grt{} algorithm. Furthermore, we show using simulations that the gap resulting from the present sorting-based algorithm decreases exponentially with increasing $n$. Moreover, the time overhead due to sorting is negligible, which makes the sorting-based algorithm also practically useful. \\

This paper is structured as follows: Section~\ref{sec:notation} introduces the notation. In section~\ref{sec:algos} we introduce two sorting-based algorithms: \sgt{} and a distribution-based sorting algorithm. We investigate their theoretical time complexities and compare the results with \grt{} in Section \ref{sec:results}. Finally, Section \ref{sec:discuss} concludes the paper with a discussion and notes on future work.

\section{Notation}\label{sec:notation}
We are given a set of $n$ weighted balls $W_i$ and $m$ bins. The total weight of a bin $U_i$ is given by the sum of the weights of the balls it contains after all balls are assigned. Since we know the ball weights \textit{a priori}, we can easily compute the ideal (but impossible since the balls are indivisible) total weight of each bin: $W_t/m$, where $W_t=\sum_{i=1}^{n}W_i$. The task is to place all $n$ balls sequentially into the $m$ bins such that the \textit{gap} $G=\max_i U_i - \min_i U_i$ is minimized. We make no assumption about the distribution from which the balls sample their weights, unless otherwise mentioned. The standard deviation of \textit{gap} is denoted by $\sigma$. \\

\section{Algorithms}
\label{sec:algos}
The goal is to minimize the gap in the $m$-bin case, where $m\geq2$. Below we use two greedy algorithms to approximately solve this problem. We compare them with each other both theoretically and in numerical experiments.

\subsection{\grt{}algorithm}
The online version of the \grt{}algorithm has previously been proposed~\cite{Azar:1994,Azar:1999} and extended to the weighted balls case~\cite{Talwar:2007}. Talwar \textit{et al.}~have shown that the weight difference between the average and heaviest bin is independent of $n$.  The only modification to the problem in the offline version is that we are given the $n$ balls \textit{a priori}. The algorithm chooses a ball $W_i$ and places it into the lightest bin.  Ties are broken arbitrarily. The algorithm is repeated until all balls have been assigned. The time complexity of this algorithm is $\Theta(n)$, since we go through all balls exactly once. The pseudo-code of this algorithm is given in Algorithm \ref{code:gr}. \\

\begin{pseudocode}[framebox]{Greedy[m]}{U_{1...m},W}
	\COMMENT{Given are a set $W$ of $n$ balls and the bin arrays $U_{1...m}$} \\
   	\COMMENT{Assign the first value to the first bin} \\
	U_1[1] \GETS W[1] \\
	\COMMENT{Initialize the pointers for all bins} \\
	p[2...m] \GETS 1 \\
	\COMMENT{First bin has already one ball in it.} \\
	p[1] \GETS 2 \\
	
	\COMMENT{Give remaining $n-1$ balls sequentially to lightest bin} \\
   	\FOR i \GETS 2 \TO n \DO
		\BEGIN 
			\COMMENT{Find the ID of the lightest bin which is the one with least current sum} \\
			idx \GETS findLightestBin(U_{1...m}) \\
			U_{idx}[p_{idx}] \GETS W[i] \\
			p_{idx} \GETS p_{idx} + 1 \\			
		\END \\
	\RETURN{U_{1...m}}
	\label{code:gr}
  \end{pseudocode}
\subsection{Sorting-based algorithms}

Sorting-based algorithms consist of two phases: sorting and greedy placement. The latter then amounts to applying \gr{} to balls sorted in the order of descending weights, such that $W_1\geq W_2 \geq, \ldots, W_{n-1} \geq W_n$. Starting from $W_1$ all balls are thrown sequentially into the bin with least current total weight. \\

In addition to finding the best-balanced allocation, it is also important to devise practically usable sorting-based algorithms. This can be accomplished by exploiting any given information about the problem. For instance, depending on the available knowledge about the weight distribution, we can propose two different sorting strategies. Regardless of which sorting strategy is chosen and which sorting algorithm is used, however, the resulting \textit{gap} is the same for all sorting-based algorithms. The pseudo-code of a sorting-based algorithm called \sgt{} is shown in Algorithm~\ref{code:sg}. \\

\begin{pseudocode}[framebox]{SortedGreedy[m]}{U_{1...m},W}
	\COMMENT{Given are a set $W$ of $n$ balls, and the bin arrays $U_{1...m}$} \\
   	\COMMENT{Sort the array in descending order (e.g. using quicksort)} \\
   	sortedW \GETS quicksort(W)  \\
	\RETURN{\CALL{Greedy[m]}{U_{1...m},sortedW}} \\
	\label{code:sg}
  \end{pseudocode}
  
\subsection{Uniform weight distribution}

If the weights are sampled from a uniform distribution over the interval $[0,A], A \in \mathbb{R}^+$, we can use a distribution-based sorting algorithm, such as bucketsort, Proxmap-sort~\cite{Standish:1997}, or flashsort~\cite{Neubert:1998}. Since these algorithms are not comparison-based, the $\Omega(n \log{n})$ lower bound for comparison-based sorting does not apply to them. For example, Proxmap-sort~\cite{Standish:1997} has an average time complexity of $\BigO{nk}=\BigO{n}$, where $k<n$ is the content number of ``buckets" used for sorting. 
Thus, the algorithm outperforms the lower bound for comparison-based sorting for large $n$. The worst-case complexity of distribution-based sorting algorithms, however, is $\BigO{n^2}$ as $n$ approaches $k$. 
However, the probability of the worst case scenario (i.e., having $k=n$ buckets) is small since $k$ is user-defined. For flashsort $k=0.42n$ has been found a good value in empirical tests~\cite{Neubert:1998}.  

\subsection{Other distributions}
For non-uniform weight distributions, we resort to efficient comparison-based sorting algorithms, such as mergesort or quicksort~\cite{Hoare:1962}), which have an average time complexity in $\BigO{n \log{n}}$. Depending on the specific sorting algorithm, the worst-case complexity can also be in 
$\BigO{n \log{n}}$. Highly optimized implementations of these algorithms are commonly available, rendering them useful in practice.

\section{Simulation results}
\label{sec:results}

We implement both \grt{} and \sgt{} in MATLAB (R2012a, The Mathworks, Inc., Natick, MA, USA). \sgt{} uses MATLAB's intrinsic quicksort function to sort the balls according to their weights. The balls are assigned random weights sampled from a uniform distribution over the interval $[0,10]$. Each simulation is repeated 1000 times with different random weights, and we report the mean and standard deviation of the gap for different numbers of balls and bins. 

\subsection{Increasing $n$}

\begin{figure}
\centering
	\begin{subfigure}[b]{0.49\textwidth}	
	      	\includegraphics[width=\textwidth]{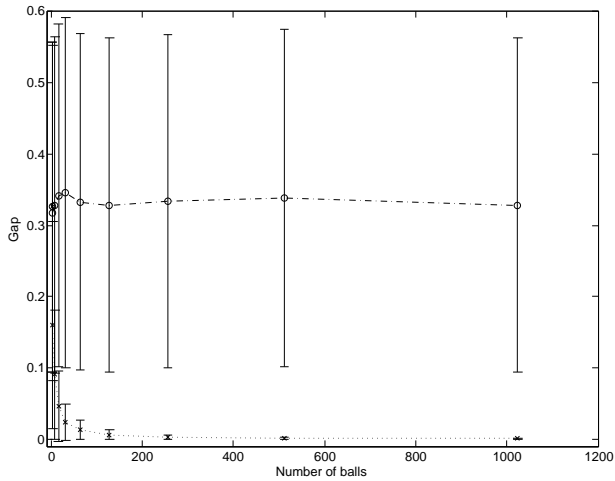} \\
		\centering (a) $m=2$
	\end{subfigure} 
	\begin{subfigure}[b]{0.49\textwidth}	
	      	\includegraphics[width=\textwidth]{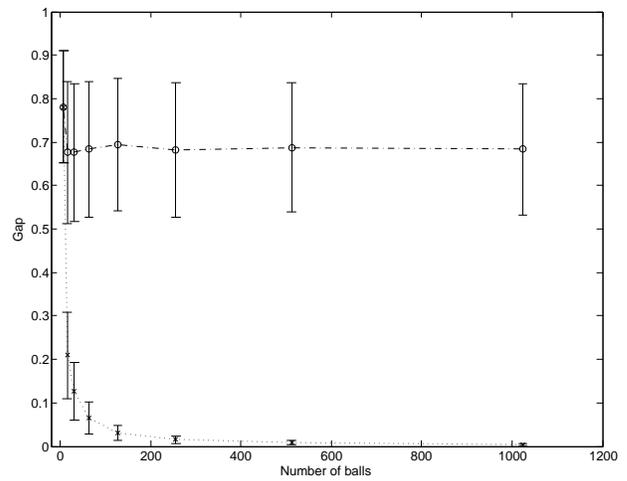} \\
		\centering (b) $m=8$
	\end{subfigure} 
	\caption{The \textit{gap} is shown for each $n$.  On average, the gaps achieved by \sgt{}(`$\circ$') are an order of magnitude smaller than those obtained by \grt{}(`$\triangle$'). (a) The case for $m=2$ bins. For $n\geq32$, the average \textit{gap} ratio between the two algorithms increases to 60. (b) The case for $m=8$ bins. Here, the \textit{gap} ratio is about 73 for $n\ge 512$.}
	\label{fig:gaps}
\end{figure}

Figure~\ref{fig:gaps} shows the results for $m=\{2,8\}$ bins and varying numbers of balls. The $\sigma$ bars for \grt{} are independent of $n$ with $\sigma=0.23$ for $m=2$ and $\sigma=0.15$ for $m=8$. 
For \sgt{} the average $\sigma$ is 0.01 for $m=2$ and 0.03 for $m=8$. \\

As seen in Fig.~\ref{fig:gaps}, \sgt{} outperforms \grt{} in all tested cases, including those with odd numbers of balls. The \textit{gap} resulting from \sgt{} decreases exponentially as the number of balls increases, and it is at least 10 times smaller than the gaps obtained by \grt{} when $n\gg m$. For each $m$-bin problem, the standard deviation across the random repetitions of the \grt{} algorithm remains constant. Also, the \textit{gap} resulting from \grt{} remains almost constant with $n$. \\

\subsection{Increasing $m$}

\begin{figure}
\centering
	\begin{subfigure}[b]{0.49\textwidth}	
	      	\includegraphics[width=\textwidth]{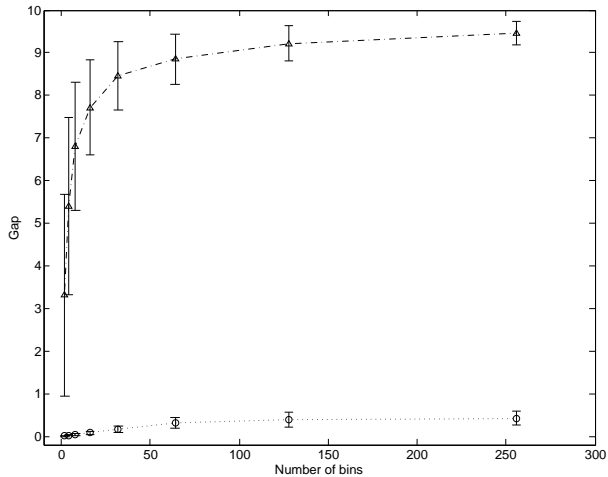} \\
		\centering (a) $n=1024$
	\end{subfigure} 
	\begin{subfigure}[b]{0.49\textwidth}	
	      	\includegraphics[width=\textwidth]{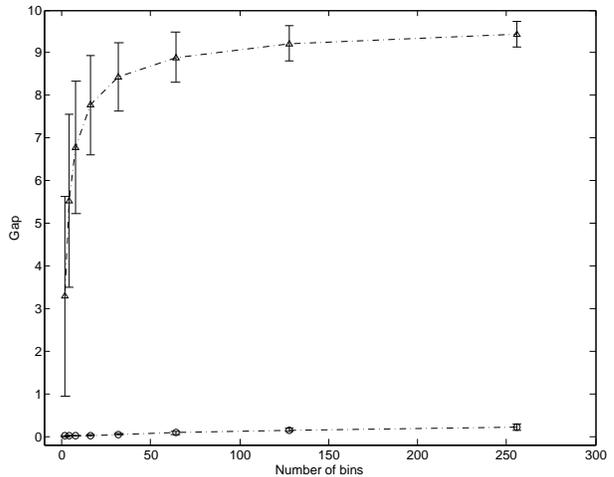} \\
		\centering (b) $n=3027$
	\end{subfigure} 
	\caption{The \textit{gap} achieved for different numbers of bins and a constant number of balls: (a) 1024 balls, (b) 3027 balls. The results are shown for the \sg{} algorithm (`$\circ$') and the \gr{} algorithm (`$\triangle$').} 
	\label{fig:gaps_bins}
\end{figure}

 In Fig.~\ref{fig:gaps_bins} we show the dependence of the \textit{gap} on the number of bins $m$ for $n=\{1024,3027\}$. The gap obtained by \gr{} first increases rapidly and then seems to saturate. That from \sg{} initially increases much slower. This is in line with previous findings~\cite{Talwar:2007}. Indeed, Talwar \textit{et al.}~\cite{Talwar:2007} show that the gap depends on both the distribution from which the weights are sampled, and on $m$.\\

\subsection{Timings}

We perform runtime measurements for the two-bin problem with $n=2^{13}$. The experiment is repeated 100 times and averages are recorded. All test runs are conducted on a Macbook Pro (MacOS X 10.7.5) with a quad-core 2.3\,GHz Intel Core i7 processor and 8\,GB 1600\,Mhz DDR3 memory. Both algorithms require approximately the same time to solve the two-bin problem. For placing $2^{13}$ balls 0.1950\,s are needed by \sgt{} and 0.1948\,s by \gr{}. Thus, sorting adds an overhead of about 2\,ms, which is $0.02\%$ of the total runtime. Increasing $m$ has no substantial effect on the final runtime as long as $n\gg m$. \\

\section{Discussion}\label{sec:discuss}
We outlined two algorithms, \gr{} and \sg{}, to solve the offline version of the weighted balls-into-bins problem. We compared their asymptotic time complexities and simulation performances.  \sg{} finds at least an order of magnitude better \textit{gaps} compared to \gr{} for $m\leq32$ and $n\gg m$. The difference grows with increasing problem size. For $n\geq4096$ the \textit{gap} ratio is at least two orders of magnitude in favor of \sg{}.  \\

The \textit{gap} from \sg{} decreases exponentially with increasing numbers of balls. Moreover, \sg{} is only weakly affected by increasing numbers of bins. The time complexity of \gr{} is in $\BigO{n}$. When the balls sample their weights from an uniform probability distribution, \sg{} has the same asymptotic time complexity. For other weight distributions, the runtime of \sg{} is in $\BigO{n \log{n}}$. As shown by our numerical experiments, however, this only incurs a minor toll in practice and both algorithms execute in almost identical times. Therefore, we conclude that the \sg{} algorithm is favorable for approximately solving instances of the offline weighted balls-into-bins problem in practice. \\
 
Future work will consider the design of a distributed dynamic load balancing protocol based on \sg{}. Such a protocol could be used in high-performance computing system for more efficiently solving task-to-processor assignment problems arising in real-world applications. 

\section*{Acknowledgements}
We thank all members of the MOSAIC Group (MPI-CBG, Dresden) and Dr. Erdem Y\"or\"uk (Johns Hopkins University) for many fruitful discussions.
 
\newpage
\bibliography{SGBIB}
\end{document}